\documentclass[journal,comsoc]{IEEEtran}

\usepackage[T1]{fontenc}% optional T1 font encoding
\usepackage{amsmath}
\usepackage{graphicx}
\usepackage{url}
\usepackage[cmintegrals]{newtxmath}
\begin{document}
\title{6G Communication Technology: A Vision on Intelligent Healthcare}

\author{Sabuzima Nayak and Ripon Patgiri,~\IEEEmembership{Senior Member, IEEE}%
\thanks{Sabuzima Nayak and Ripon Patgiri, Department of Computer Science \& Engineering, National Institute of Technology Silchar, Cachar-788010, Assam, India}% <-this % stops a space
\thanks{Email: sabuzimanayak@gmail.com and ripon@cse.nits.ac.in (URL: http://cs.nits.ac.in/rp/)}
\thanks{Corresponding author: Ripon Patgiri}
\thanks{Manuscript received Month 00, 20XX; revised Month 00, 20XX.}}%

\markboth{IEEE Internet of Things Journal,~Vol.~00, No.~00, Month~20XX}%
{{Nayak and Patgiri}: 6G Communication Technology: A Vision on Intelligent Healthcare}
\maketitle

\begin{abstract}
6G is a promising communication technology that will dominate the entire health market from 2030 onward. It will dominate not only health sector but also diverse sectors. It is expected that 6G will revolutionize many sectors including healthcare. Healthcare will be fully AI-driven and dependent on 6G communication technology, which will change our perception of lifestyle. Currently, time and space are the key barriers to health care and 6G will be able to overcome these barriers. Also, 6G will be proven as a game changing technology for healthcare. Therefore, in this perspective, we envision healthcare system for the era of 6G communication technology. Also, various new methodologies have to be introduced to enhance our lifestyle, which is addressed in this perspective, including Quality of Life (QoL), Intelligent Wearable Devices (IWD), Intelligent Internet of Medical Things (IIoMT), Hospital-to-Home (H2H) services, and new business model. In addition, we expose the role of 6G communication technology in telesurgery, Epidemic and Pandemic.
\end{abstract}

\IEEEpeerreviewmaketitle
\section{Introduction}
\IEEEPARstart{6}{G} communication technology is attracting many researchers due to its prominent features and its promises. It will revolutionize diverse fields and we will evidence the revolution from 2030 onward. Many features of 6G have already been discussed in premier forum and continuously gathering various requirements of 6G communication technology \cite{Saad,Gui,Giordani,Chen}. Also, Nayak and Patgiri \cite{Nayak} exposes issues and challenges of 6G communication technology. Many countries have already started 6G communication technology for timely deployment. Firstly, Finland initiated 6G project in 2018\cite{Katz}. Secondly, United State, South Korea and China have started 6G project in 2019 \cite{Dang}. Recently, Japan has also initiated 6G research project in 2020 \cite{Docomo}. In addition, many algorithms have been developed for 6G \cite{Dong,Mao}. Now, it is essential to initiate 6G project to not be left behind by other countries. On the contrary, 5G communication technology is yet to deploy in full scale over worldwide and B5G yet to be developed. 5G and B5G will have various drawbacks for revolutionize modern lifestyle, society and business. For instance, unable to support holographic communication due to lower data rate. Therefore, it is the peak time to envision future possibilities of 6G communication technology. Also, it is necessary to envision the future healthcare for well-being of peoples of the society. 

The current healthcare system is providing basic facilities and the key barrier of the current healthcare system is time and space. This is unavoidable in the current scenario, however, it will not be a barrier in the near future. Moreover, ambulance service is just a transporter of patients with oxygen facility and road traffic priority which can be served by a normal car too. Besides, the elderly service is very unsatisfactory in current scenarios. The elderly service requires intensive care from medical staffs. However, it is unavailable till date. Most of patient dies in ambulance while travelling from home to hospital or before ambulance reaching the spot. Also, the accident detection system is unavailable in current healthcare systems. The accident detection system requires real-time detection to provide medical services on time and on the spot. Furthermore, Epidemic and Pandemic outbreaks, for instance, COVID-19, cannot be controlled due to lack of advanced infrastructure. A similar kind of virus will again arise in future. Thus, it is utmost important to develop the intelligent healthcare system.

\begin{figure*}[!ht]
    \centering
    \includegraphics[width=\textwidth]{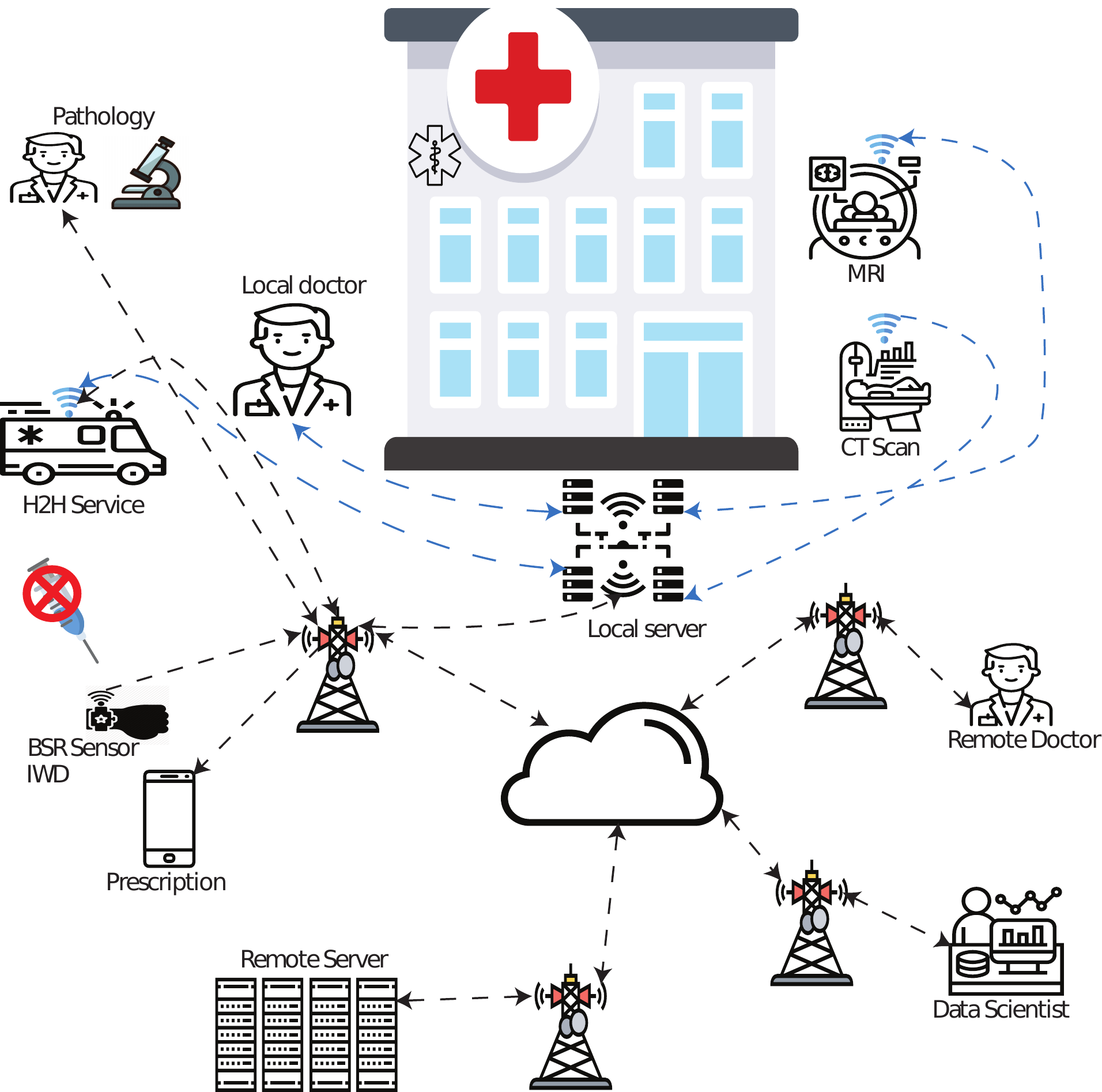}
    \caption{\textbf{A landscape of intelligent healthcare systems. This figure includes Intelligent Internet of Medical Things (Blood Sample Reader (BSR) sensor, Intelligent Wearable Devices (IWD), Online Prescription, MRI, CT Scan), Hospital-to-Home (H2H) Services implements mobile hospital, Pathology, Local Doctors, Remote Doctors, and Data Scientist.}}
    \label{fig1}
\end{figure*}

The requirements of 6G communication technology for future healthcare are high data rate ($\geq 1~Tbps$), high operating frequency ($\geq 1~THz$), low end-to-end delay ($\leq 1~ms$), high reliability ($10^{-9}$), high mobility ($\geq 1000~km/h$) and wavelength of $\leq 300 \mu m$ \cite{Chen}. In particular, telesurgery requires real-time communications. Also, holographic communication and augmented/virtual reality will boost up the intelligent healthcare systems. However, 5G and B5G will be unable to support intelligent healthcare. In the 5G communication era, intelligent healthcare will be implemented partially which will push forward a step ahead. However, the rural connectivity is still a challenge for 5G communication technology \cite{Yaacoub} as well as healthcare. Communication technology boost up the performance of health service and a landscape of intelligent health service is depicted in Figure \ref{fig1}. 

It is expected that 6G communication technology will revolutionize the healthcare completely and the healthcare will fully depend on communication technology. We will evidence the paradigm shift in healthcare due to the advent of communication technology. Current state-of-the-art healthcare system is unable to provide telesurgery due to communication issues. Moreover, ambulance service is to be replaced. Wearable devices need to be redefined. The hospital is required to restructure. The health services should be provided in real-time. Health monitoring and elderly services need to be redefined. Therefore, in this perspective, we envision the future healthcare using 6G communication technology. We brief the required parameters of 6G communication technology and its prime technologies. Also, we focus on Quality of Services (QoS), Quality of Experiences (QoE) and Quality of Life (QoL). Moreover, we envision the requirement of Hospital-to-Home (H2H) services, Blood Sample Reader (BSR) sensor, Intelligent Wearable Devices (IWD), and Health Insurance at Hospital (HIH) business model. Moreover, the role of Artificial Intelligence (AI) and Edge technology in healthcare is exposed. Thus, Section \ref{tech} surveys detail parameters of 6G communication technology to establish the understanding of the article. In addition, Section \ref{ena} reviews the enabling technology for 6G to meet intelligent healthcare in near future. Sections \ref{hol}, \ref{aug} and \ref{tac} expose the role of holographic communication, augmented and virtual reality, and tactile Internet in intelligent healthcare. Moreover, the parameters of intelligent healthcare are envisioned in Sections \ref{imt}, \ref{h2h} and \ref{tel}. This article also exposes the role of communication in handling Epidemic and Pandemic in Section \ref{epi}. Furthermore, A new business model also discussed in Section \ref{bus}. Most importantly, intelligent healthcare requires security, secrecy and privacy which are addressed in Section \ref{sec}. Finally, this article is concluded in Section \ref{con}.

\section{6G Technology} % Done 
\label{tech}
6G will be using the terahertz (THz) signal for transmission \cite{Gui}. THz signal increases bandwidth and data rate. Moreover, it will provide bandwidth three times higher than 5G signal, i.e., mmWave \cite{Chen1}. 6G will have a data rate of 1 TBPS or more. 5G and B5G follow 2-dimensional  communication structure, whereas the 6G will follow 3-dimensional, i.e, time, space and frequency. 6G will provide the 3D services with the support of emerging technology such as edge technology, AI, cloud computing and blockchain. The 6G communication network will be ubiquitous and integrated \cite{Al-Eryani}. 6G will provide deeper and broader coverage through device to device, terrestrial and satellite communication. 6G aims to merge computation, navigation and sensing to the communication network. In the area of security, 6G will cover security, secrecy and privacy of the Big data generated by billions of smart devices. However, there will be a transition from smart device to intelligent device. 

\subsection{Requirements} 
The requirements of 6G communication are gathered and discussed by various researchers \cite{Katz1,Piran,Zong,Zhang,Zhang1,Zhang2,Zhang4,Tomkos}. The key requirements of 6G communication technology are 1 THz operating frequency, 1 Tbps data rate, 300 $\mu m$ wavelength, and 1000 kmh mobility range \cite{Saad,Nayak}. The 6G architecture is 3D with consideration of time, space and frequency. The end-to-end delay, radio delay and processing requirements are $\leq$ 1 ms, $\leq$ 10 ns and $\leq$ 10 ns respectively to provide real-time communication \cite{Gui,Rappaport}. Moreover, 6G will be truly AI-driven communication technology \cite{Nayak}. It is expected that 6G will be fully backed by satellite. The NR-Lite will be replaced by Intelligent Radio (IR) \cite{Letaief}. Also, the core network the Internet of Things (IoT) will be replaced by the Internet of Everything (IoE). It is expected that 6G will enable and revolutionize many technologies in coming future. We will evidence the transitions of IoT to IoE, Smart Devices to Intelligent Devices, and other numerous possibilities.

\subsection{Terahertz Communication}
6G will use terahertz radiation (THz) also called submillimeter radiation. It is electromagnetic waves with wavelengths in between 1mm to 0.1 mm, hence called submillimeter radiation. THz has many advantages making it suitable for 6G communication such as high bandwidth, high data rate (upto several Gbps), high capacity and high throughput \cite{Chen}. Efficiency of THz signals is possible to increase by spectrum reuse and sharing. Some techniques already exist for spectrum reuse such as cognitive radio (CR). It helps many wireless systems to access the same spectrum through spectrum sensing and interference management mechanisms \cite{Chen}. In case of spectrum sharing, temporally underutilized or unlicensed spectrum is utilized to maintain availability and reliability. Symbiotic radio (SR) is a new technique to support intelligent, heterogeneous wireless networks. It will help in efficient spectrum sharing. However, deploying these techniques in the 6G wireless network are still big challenges. Moreover, continuous THz signal generation is strenuous due to requirement constraints regarding size. Designing an antenna/transmitter is also complex. In addition, the signal gets attenuated to zero after short distance transmission. It occurs due to energy loss, i.e., molecular absorption and spreading loss \cite{Han}.

\subsection{Transition from Smart to Intelligent}
6G is expected to be truly AI-driven communication technology and hence, it will be able to introduce intelligent era \cite{Nayak}. Therefore, all smart devices will be converted to intelligent devices in the era of 6G \cite{Nayak1}. The advent of AI along with mobile communication, we will evidence many transitions from smart things to intelligent things. From 2030 and onward, the IoT will be intelligent and will be replaced by IoE. The smart phone will be replaced by intelligent phones. Intelligent devices will be AI-driven devices that are able to connect to the Internet. Thus, the intelligent device (may be tiny device) will be able to predict, make a decision and share their experience with other intelligent devices. Therefore, there is a paradigm shift from smart to intelligent era using 6G communication technology.

\subsection{Quality of Services}
The Quality of Services (QoS) parameters of 6G technology are higher than 5G and B5G. For instance, QoS includes high data rates, extremely reliable and low latency communication (ERLLC), further-enhanced mobile broadband (FeMBB), ultra-massive machine-type communications (umMTC), long-distance and high-mobility communications (LDHMC) and extremely low-power communications (ELPC) \cite{Zhang1}. Also, QoS includes mobile broad bandwidth and low latency (MBBLL), massive broad bandwidth machine type (mBBMT), massive low latency machine type (mLLMT) \cite{Gui}. This QoS service parameters enables diverse applications to revolutionize. Thus, the 6G technology will be proven as a revolutionary technology in many fields along with healthcare.

\subsection{Quality of Experiences}
The Quality of Experiences (QoE) defines a high QoS and user-centric communications. QoE will be achieved by holographic communications, augmented reality, virtual reality, and tactile Internet which requires high data rate with extremely low latency. Moreover, QoE is expected to be revolutionary in intelligent cars, intelligent devices, intelligent healthcare, intelligent drones and many more. A high QoE can be achieved only when all desired parameters are implemented by 6G technology. 6G technology will be truly AI-driven communication technology \cite{Nayak}, and thus, we will evidence many changes in our lifestyles, societies and businesses. Also, 6G promises to provide five sense communication to provide rich QoE. Thus, 6G communication will be a major milestone for healthcare. Healthcare requires high QoE for critical operations, intelligent hospitals and intelligent healthcare.

\subsection{Quality of Life}
The Quality of Life (QoL) is defined as enhanced lifestyle with QoS and QoE in healthcare. 6G technology will enable high QoL using communication technology. The QoL is not a core parameters of 6G communication technology, however, it will be core parameters of the intelligent healthcare systems. 6G will be able to provide high QoE along with desired parameters of QoL. The key parameters of QoL are remote health monitoring of patients, including elderly persons, connection to intelligent wearable devices (IWD), intelligent accident detection, telesurgery, precision medicine, and many parameters will be included in future. Also, another prominent feature of QoL is Hospital-to-Home (H2H) services. H2H service will be implemented using mobile hospital in an intelligent vehicle. The mobile hospital will have minimum requirements to be a hospital including medical staffs. This service is essential in QoL to improve the modern lifestyle and emergency services. In addition, Health Insurance at Hospital (HIH) will emerge to replace the current health insurance policy. HIH will increase faith on hospital services and it is also included in QoL. The Hospitals will be ranked based on QoL parameters in near future.

\section{Enabling Technology for 6G}
\label{ena}
6G communication technology requires supporting technologies to fulfill the promises. 6G is truly AI-driven communication technology \cite{Nayak}, and thus, it requires AI to integrate its communication technology. Moreover, 6G will enable Internet of Everything (IoE), and it will boost up many fields. Also, edge technology is necessary for 6G technology for bringing the Cloud features closer to intelligent devices. Thus, 6G communication technology comprises of many technologies.

\subsection{Internet of Everything (IoE)} %Done for healthcare
6G follow 6C’s for communication \cite{Gui}. 6C’s are capture, communicate, cache, cognition, compute, and control. High level sensing is capturing. It is essential for holographic communication in healthcare. The captured data are converted to digital data, stored in a local cache and transmitted to remote locations in real-time. In some cases, digital data are further converted to signals and transmitted to other devices for processing. But before computing, cognition helps in formulating feasible determinations based on input digital data. These are intelligent determinations which helps in making the computing easy. The computed data are transmitted to smart devices to help in controlling the action taken by smart devices for healthcare \cite{Gui}. For example, raising an alarm on Epidemic and Pandemic. It will use the core services of combined and enhanced eMBB and mMTC. It will require high data rates to support touch experiences in the intelligent devices for healthcare. IoE will be required by 6G to have a huge capacity to connect millions of intelligent devices, collect tactile sensations and convert to digital data \cite{Gui}. Huge capacity to connect the sensors and actuators in healthcare communication. And, low latency to maintain seamless integration among them \cite{Gui}. When 6G will be commercial, it will not be the era of Big data but rather Big data 2.0. Big data 2.0 require a supercomputer to compute and analyze massive scale of small-sized data produced by healthcare devices \cite{Nayak1}. 

\subsection{Edge Intelligence} %Done.
6G will rely on Cloud computing for storage, computing and analysis of Big data \cite{Nayak}. Data produced by the intelligent devices are transferred to Cloud for storage, however, it consumes communication resources and bandwidth. Nowadays, the technologies are brought closer to the data source due to the exponential growth of data. This technology is Edge technology. 6G is claiming to have a high capacity to provide smooth services to billions of intelligent devices \cite{ullah}.  6G will rely on Edge technology to provide the smooth and high speed Internet services to the intelligent devices which is vital for healthcare. Edge technology collects, computes and analyses the health data in real time in its Edge nodes \cite{Mao}. These nodes are located closer to the intelligent medical devices. The data generated by the user are transmitted to the Edge nodes. Also, the Edge nodes compute the health data. Then, Edge nodes analyze the data to decide the appropriate action. For example, Edge node will receive user health data transmitted from the intelligent medical devices and determines whether the user is suffering from any deficiency. The health data are constantly transmitted to Edge nodes. The Edge node monitors the health related data. Edge node also filters the health related data and transmits the important information to the Cloud for storage. Edge technology reduces cost of communication and computation \cite{Illa}. Some other advantages of Edge technology are low latency, reliability, privacy, scalability and adaptability. The massive number of intelligent devices will connect to 6G Internet, hence, all the advantages of Edge technology will greatly help 6G to meet its requirements to provide high QoS in healthcare. 

% \begin{figure}[!ht]
%     \centering
%     \includegraphics[width=0.5\textwidth]{edge.pdf}
%     \caption{\textbf{Sub-network of 6G communication using Edge technology.}}
%     \label{fig2}
% \end{figure}

“The marriage of edge computing and AI gives the birth of Edge intelligence \cite{Zhou}.” Edge intelligence is implemented using AI algorithms for analysis in Edge nodes called Edge analytics. Edge nodes receive a huge volume of health related data and AI is used to find patterns or compute them for analysis. With the help of AI, Edge nodes are capable of developing image, data and video Edge analytics. GigaSight \cite{GigaSight} is a video Edge analytics with a decentralized hybrid cloud architecture. Xie \textit{et al.} \cite{Xie1} proposed a video analytics that has lightweight virtualization by implementing container technology. For filtering of data in Edge nodes, Nikolaou \textit{et al.} \cite{Nikolaou} proposed a predictive Edge analytics. Similarly, Cao \textit{et al.} \cite{Cao} proposed a descriptive analytics for mobile Edge nodes. However, AI algorithms are computationally intensive. Currently, AI algorithms execution requires high computational resources, and power consumption, which is limited in the Edge nodes. Hence, they depend on the Cloud for execution. However, Edge nodes will be one of the network nodes in 6G. All 6G network nodes will be AI-enabled for providing intelligence services to the healthcare systems. Moreover, 6G will have real-time intelligent Edge which will dramatically boost up the healthcare system. It will perform computing and analysis on live data \cite{Huang}. Therefore, it is very important to execute the AI algorithms in Edge nodes instead of Cloud to reduce latency in providing the services \cite{Al-Eryani}.

\subsection{Artificial Intelligence} %Done
6G will be a truly AI-driven communication network\cite{Nayak,Nawaz}. 6G will make every aspect of network communication intelligent to make the system self-aware, self-compute and self-decide on a situation. The goal of 6G is to provide global coverage, including space-air-water. This is achievable only by making the different aspects of communication ``intelligent”. Implementation of AI algorithms is generating high accuracy and performance in communication networks. Truly AI-driven communication can offer real-time communication which is very important for modern healthcare.

AI-driven healthcare improves clinical diagnosis and decision making \cite{Yu}. The healthcare requires AI to perform tasks in real-time. Deep learning (DL) does not require data preprocessing. It takes original health data and performs the computation, thus, real-time data can be given as input. Moreover, it shows high accuracy while computing a large number of network parameters\cite{Mao1}. Similarly, another AI algorithm that is currently explored on health data is Deep Reinforcement Learning (DRL). In reinforcement learning, the system first develops a few decisions, then observes the results. Based on the observation the decision is again computed to obtain an optimal decision. DRL combines both reinforcement learning and deep neural networks algorithms and combines the advantages of both \cite{Luong}. Thus, DRL gives high performance within small computation time. In addition, federated AI will share their knowledge among the intelligent devices which will booster healthcare. AI algorithms have shown high performance. AI algorithms require expensive infrastructure. The AI is also preferred for proactive caching. For Big Data, parallelism in training should be explored. All AI algorithms perform high computation task. The high computation task takes a long time and consumes more power. Whereas, 6G is unable to provide such relaxation. The AI algorithms that will be implemented in 6G will have their own issues. For example, large numbers of layers in Neural Network. However, research is performed to improved AI algorithms with less computation time and less energy consumption to improve the performance of 6G which leads to increase in efficiency in healthcare.

%%%%%%%%%%%%%%%%%%%%%%%%
\section{Holographic Communication} %Done->Very Good.
\label{hol}
“Hologram is a physical recording of an interference pattern that uses diffraction to generate a 3D light field \cite{holo}”. The image generated has parallax, depth and other properties of the original object. Holographic communication uses cameras from different angles to create a hologram of the object. It will use the core service of combined and enhanced eMBB and URLLC. It will require high data rates to provide good quality of service and streaming high definition videos. Moreover, very low latency is required for real-time voices and immediate control responses \cite{Gui}. Holographic communication will be a major breakthrough for healthcare. And, 6G is  capable of providing this service. 6G Holographic communication will help connect people. In case of emergence, in some cases the doctors have to wait for the expert doctor(s) for a diagnosis. However, using holographic communication the expert doctor(s) can diagnose the patient while travelling. And, can supervise the doctor for an early medical treatment. Many times the patient has to travel to many doctors for a correct diagnosis or the treatment is unavailable in that hospital. The patient may have to travel to different states or to different countries. In such cases, it becomes an economical and physical burden for the patient. Also, travelling in bad health is very stressful for the patient. However, using holographic communication the doctors can diagnose remotely. The patient can only visit the hospital for treatment. Holographic communication will also help the expert doctors provide services in rural areas while staying in cities or towns. Similar to 6G global coverage, the 6G holographic communication will help in global connectivity of healthcare. Upon request the expert doctor can instantly provide services without having to adjust their schedule and travel. Moreover, it will be possible to connect different doctors around the world to discuss/supervise in complex medical cases. 

\section{Augmented Reality and Virtual Reality}%Done->Excellent
\label{aug}
Augmented reality (AR) helps to include virtuality to real objects. Moreover, it is combined with multiple sensory abilities such as audio, visual, somatosensory, haptic etc. AR also provides real-time interaction, and presents 3D images of virtual and real objects accurately. Virtual reality (VR) refers to presenting an imaginative or virtual world where nothing is real. AR and VR will use the core service of combined and enhanced eMBB and URLLC. It will require high data rates to provide good quality of service. Moreover, to stream high definition videos. Moreover, very low latency is required for real-time voices and immediate control responses \cite{Gui}. AR and VR require a peak data rate of 1 Tbps and user experience of $>$10 Gbps with $>$0.1 ms latency which can be provided by MBBLL\cite{Gui}. Currently, both AR and VR are developing. However, 6G will open new windows for their usage in the field of healthcare. AR will help to view the inside of the body of a patient without any incision. Moreover, doctors can adjust the depth of the specific location in the body \cite{Blum}. The specific body area can also be enlarged for better visibility. 6G will help the doctors to view a patient remotely. The AR and holographic communication can be combined for better diagnosis. Using VR the doctors can practice medical procedures without any patient. It will be very helpful in case of practicing complex procedures/surgery having high risk. All these devices will be intelligent devices and connected to 6G Internet. As discussed above using 6G a smooth and high resolution presentation can be created for remote medical learning or diagnosis. 

\section{Tactile/Haptic Internet} %Done->Excellent
\label{tac}
Haptic technology creates a virtual touch using force, motion or vibration on the user. Tactile Internet is used to transfer the virtual touch to another user, maybe human or a robot. Tactile Internet requires high speed of communication and ERLLC to grab the tactile in real-time. This technology will be used for remote surgery, i.e., telesurgery. It will also help doctors for diagnosis using touch without being physically present. Haptic human-computer interaction (HCI) is classified into three types, namely, desktop, surface, and wearable \cite{Haptic}. In desktop HCI, the remote doctor will be able to use a virtual tool for surgery or diagnosis. In surface HCI, the movement is not 3D, but 2D.  The device to give command has a flat screen such as a mobile or tablet. As moving the hand on the screen the robot can be given a command to interact with the patient. In wearable HCI, for instance a haptic glove, is used by the remote doctor. Tactile/haptic technology will also help in providing healthcare during disaster time. For example, the COVID-19 when all countries are under lockdown and interaction with outside of the state and country is closed or natural disasters. During such situations, tactile/haptic technology will help in healthcare. Expert doctors can perform complex surgery remotely using robots. In addition, during epidemics and pandemics, the medical personnel are at great risk. A little carelessness may expose them to the deadly/contagious disease. Using tactile/haptic technology robots can be used to interact or care for the patient while medical personnel will be present remotely. All this is achievable by using 6G high speed and low latency Internet. 

%%%%%%%%%%%%%%%%%%%%%%%%%%%%%%%%%%%%%%%%%%%%%%%%%%%%%%
\section{Intelligent Internet of Medical Things}
\label{imt}
In 6G communication paradigm, Intelligent Internet of Medical Things (IIoMT) will evolve and serve many purposes for well-being of humankind. IIoMT are intelligent devices that are AI-driven that makes its own decision using communication technology. IoE will also emerge along with IIoMT, and thus, medical things can connect to the Internet. For instance, MRI and CT scan. The scanner will scan the devices and send the data to remote locations through 6G technology as depicted in Figure \ref{fig1}. These data can be analyzed by a pathologist in real-time. Almost all medical things will be able to connect to the Internet and instant decision can be taken. Therefore, IIoMT will be able to overcome the barrier of time, space and money. Another example, cancer patients can easily be treated by remote doctors. Currently, it takes time to detect whether the cancer patients having benign or malignant. However, cancer can be detected in real time in the near future using 6G communication technology. Also, the doctors and patients do not have to visit specialist hospitals. It takes time and money. Therefore, remote doctors will treat the cancer patients in collaboration with local doctors. Early detection of cancer patients can reduce the mortality rate to nearly zero. However, such sensors need to be invented. This scenario not only applies to cancer, but also applies to many diseases. For instance, cardiovascular treatments.

\subsection{Blood Sample Reader} %Done
Blood is vital for the human body and most of the diseases can be detected from blood. Conventional blood sampling requires needles to inject to sample the blood from the human body. However, numerous research is being conducted for needle-free blood sampling devices. For instance, Lipani \textit{et. al.} devices new needle-free method for glucose monitoring for diabetic patients \cite{Lipani}. Blood Sample Reader (BSR) sensors will be needle-free intelligent wearable device. The BSR sensors will revolutionize the health industry. BSR will correctly read every parameters of blood, for instance, WBC, RBC, etc. This BSR sensor will remain connected to the 6G Internet. A blood sample will be transmitted to a testing center for test results periodically, automatically or with permission from the patient. Hence, no manual intervention will be required for testing Blood sample. This device will play vital role in intensive care on elderly service and monitoring. Moreover, COVID-19 outbreak could be easily stopped using BSR sensors. With the number of cases crossing millions, countries are running out of medical gears. Taking the blood sample using needle exposes the medical staff to the disease. This BSR sensor can easily track the spread of COVID-19 virus in real-time. Thus, it will be a huge help during Epidemic and Pandemic. Therefore, BSR sensor will be the most demanded medical devices in the future. 

\subsection{Intelligent Wearable Devices}
Intelligent Wearable Devices (IWD) are connected to the Internet and transmit psychological and physical data to test centers and monitoring centers. This devices will monitor heartbeat, blood pressure, blood tests, health conditions, body weight and nutrition. The test result will be received quickly. Also, IWD learn from the personal body history and advise the person for the next action, for instance, advising for walk or running. IWD will maintain a personal history of health, nutrition, and habits. Thus, IWD can advise what to eat in case of any deficiency. Detection of minor body issues such as deficiency will reduce the frequency of hospital visit. So, it will reduce hospital bill and hospital can focus on more complex diseases. In addition, IWD will read blood sample and the blood sample will be transmitted to pathological results. Thus, early detection of cancer can be possible using IWD.  Therefore, IWD can improve health conditions and increase human life span. Also, it is vital in elderly services because elderly service requires intensive cares. Future IWD will combine multiple features in a single device. All features will be packed into a single device. The initial release of such devices will be expensive, however, over a time period cost will decrease and the common people can afford. 

%%%%%%%%%%%%%%%%%%%%%%%%%%%%%%%%%%%%%
\section{Hospital-to-Home Services}
\label{h2h}
Currently, the ambulance services are just a transporter of patients with oxygen and road traffic priority. It does not serve the purpose of emergency service due to absence of intelligence. Therefore, the ambulance services are not impacting on our lives. Any normal car can also solve the same purposes if we keep oxygen and emergency signal. Therefore, a new kind of ambulance service is required to improve lifestyle. 

To replace ambulance services, the Hospital-to-Home (H2H) services will be emerging. Due to the advent of communication technology, hospital can reach to home on demand and in an emergency situation. The future vehicles will be fully AI-driven to make intelligent vehicles \cite{Tang}. Therefore, H2H will be implemented upon mobile hospital on an intelligent vehicle platform that will have a minimum dependence on hospitals including doctors and nurses. This mobile hospital will replace ambulance services. For instance, mobile hospital detects an accidents in real-time and reaches the spot. Then, the mobile hospital will start treating the patients before reaching the hospitals. Moreover, mobile hospital can detect any emergency situation in real-time and reach the spot to save lives. It will also enhance modern lifestyles. Specially, it is immensely necessary in elderly services. Thus, 6G communication revolutionizes modern lifestyle through the H2H services.

%%%%%%%%%%%%%%%%%%%%%%%%%%%%%%%%%%%%%%%%%%%%%%%%%%%%
\section{Telesurgery}
\label{tel}
The telesurgery is emerging and it is a concept for the future. It is defined as remote surgery by the doctor(s) \cite{challacombe}. Telesurgery requires robots, nurses and mediator of remote doctors. Communication plays a key role in telesurgery. Also, it requires a very high data rate and URLLC. 5G and B5G are unable to provide these requirements. Hence, telesurgery requires the support of 6G technology. Moreover, the success of telesurgery requires real-time communication. For telesurgery, the doctor can provide guidance through verbal,  telestration or tele-assist \cite{HUNG}. Due to 6G, more interactive verbal guidance can be provided using holographic communication. In holographic communication, the doctor can be present in the surgery for guidance and also can also move to have a better visual of the surgery area. Telestration is showing the surgery procedure remotely, for instance using video. In intelligent healthcare, AR and VR can be used for telestration. In addition, the doctor(s) can tele-assist the surgery using tactile/haptic technology. The requirements of telesurgery can be fulfilled by 6G communication technology and 6G will prove that surgery can be perform beyond boundaries.

%%%%%%%%%%%%%%%%%%%%%%%%%%%%%%%%%%%%%%%%%%%%%%%%%%%
\section{Epidemic and Pandemic}
\label{epi}
Communication technology will play a vital role in epidemic and pandemic. Mostly, epidemic and pandemic are human to human transmission disease. Therefore, medical staffs are at high risk and often termed as ``suicide squad''. Recently, COVID-19 has taken thousands of lives, including the medical staffs. COVID-19 outbreaks could easily be stopped using IIoMT. For instance, blood sample of each person is transmitted to testing center using BSR sensor without exposing themselves to the outer worlds and the result will be received at their own home. A blood sample will be taken by intelligent wearable technology (BSR sensor) and transmit the sample data to the test centers. Thus, human chain can easily be stopped and the outbreak can be tracked in real-time. Now, millions of COVID-19 positives have been detected and billion of testing has to be conducted. This global tragedy demands intelligent healthcare systems and IWD.

%%%%%%%%%%%%%%%%%%%%%%%%%%%%%%%%%%%%%%%%%%%%%%%%%%%%%%%%%%%%%
\section{Precision medicine}
\label{pre}
Precision medicine is developing a customized medicine or treatment for providing better treatment to a patient \cite{Nayak2}. For the development of precision medicine the doctors/researchers conduct research by grouping the people based on some common parameter. 6G technology can greatly help in the development of precision medicine. Also, it requires AI to provide personalized healthcare \cite{Reddy}. Better development of treatment requires health data of the clinical trial people. For instance, cell therapy research is carried out for critical disease treatment \cite{Scheetz,Ioannidis}. The doctors and researchers can collect the data using IWD. This data will be collected in real-time which helps in providing accurate health data. Moreover, the research can be conducted globally. Geographical condition influence a person’s immunity system. Hence, moving the people under observation to a single location will change their environment. And, this will influence the research. Therefore, through IIoMT the doctor/researcher will observe the people under observation across the globe.

\section{New Business Model}
\label{bus}
6G will revolutionize businesses, and thus, the old business model of the hospital need to be redefined. Currently, the treatment expenses are paid directly to the hospital by the patient or by the health insurance policy. The health insurance policy is operated by a company. The company forms a network of hospitals. This hospitals allows cashless treatment and incurred money is charged from the company. Most of the insurances do not cover many diseases and elderly health services. This business model defines the hospital and insurance as a different entity during treatment and payment is to be made to the hospital. The insurance company will pay the incurred money to the hospital. Excessive amount needs to be paid by the patients. Therefore, health insurance and hospital services are very unsatisfactory and they do not insure our lives. In true sense, neither hospital nor insurance company is responsible for our treatment from the date of birth to death. 
%In this case, why should people open a health insurance policy if they do not cover most of the diseases?

To that end, we envision a new business policy for modern healthcare systems. It is expected that new business model will remove health insurance companies and the hospitals will plays the role of the health insurance companies. We term it as Health Insurance at Hospital (HIH). HIH will form a network of hospitals, and remove mediator (health insurance company). HIH has to take care of geographical locations while forming network of hospitals. Yearly or monthly premium of insurance will be directly paid to the hospital. The hospital will function using the insurance money. The hospital will take care of a person's health from birth to death. A patient can call H2H service to reach the hospital. The hospital will verify the HIH identification, and then the patient will be admitted to any partner hospital. The patient will be treated with priority. After getting discharged from the hospital, the patient does not have to pay any medical bill. Also, this policy will cover QoL including IWD connection and monitoring, H2H services, telesurgery, surgery, pregnancy and delivery, vaccines, ICU services, NICU services, elderly services, personalization and precision medicines. HIH includes all types of diseases and accidents, including cancer, flu, cardiovascular, neurological problems, etc. There will be no bar in number of treatments for a patient. Also, the policy will cover OPD and all other medical expenses, including pathological and medicine expenses, that is, no bill, no claim and hassle free treatment. 

%%%%%%%%%%%%%%%%%%%%%%%%%%%%%%%%%%%%%%%%%%%%%%%%%%%%%%%%%%%%
\section{Security, Secrecy and Privacy} %Done
\label{sec}
The key focus of 6G technology is security, secrecy and privacy \cite{Dang}. Therefore, 6G requires secure URLLC (sURLLC) to have an enhanced secure communication. 6G communication technology promises the highest level of security. It will defend the attacks using federated AI, Quantum Machine Learning, Quantum Computing and THz communication. THz communication is eavesdropping and jamming proof \cite{Nayak}. Healthcare requires a high level of security for data transmission over the network. Any alteration of health data can kill a patient. Therefore, it is crucial to protect health data from attackers. Moreover, 6G communication technology also focuses on secrecy of the most sensitive data. Sensitive data are protected from anyone excepts owner of the data. Also, administrators are not permitted to view these sensitive data. For instance, family history. In addition, 6G also focuses on privacy which is crucial parameters of healthcare. To increase privacy, 6G will also rely on Edge technology. The Edge nodes are located closer to smart devices. The computed data are also analyzed in the same Edge nodes. Edge nodes have small memory, hence, all data are not concentrated in one location. Therefore, Edge maintains the privacy of the user. Another important point is filtering of data by Edge nodes. Edge nodes filter the data and transmit only important information to the Cloud. It also leads to storage of less information about the user in the Cloud. Thus, it is easier for Cloud to provide security for lesser data. In the current scenario, Blockchain provides a high degree of privacy for health data \cite{McGhin}. It provides a secure infrastructure for handling the health data \cite{Ebru}. In the future, Blockchain with more advanced techniques will be very helpful in intelligent halthcare.

\section{Conclusion} % No conclusion
\label{con}
Intelligent healthcare must enhance QoL. Precisely, intelligent healthcare comprised of IIoMT, IWD, H2H and HIH. All IIoMT requires high quality mobile communication, AI integration, and support from Edge computing and Cloud Computing. A key IIoMT is BSR sensors. BSR sensors are yet to be devised. BSR will be able to solve many issues in healthcare. Thus, BSR sensor needs extreme attention from the research community. Also, another prominent IIoMT is IWD. IWD will be a revolutionary technology for healthcare and it will be equipped with many sensors including BSR. It will help in diagnosis of many diseases automatically and improve drastically the health of a person. Also, a person does not require to visit hospital for regular check up, for instance, blood test, blood pressure, sugar level, etc. IWD sends the personal health data periodically to the health monitoring center for detection of any kind of abnormality. Also, an elderly person can be monitored without intervention of medical staffs. Elderly service requires intensive care, and the integration of IWD with the healthcare system will automatically take care of the person without intervention of medical staffs. In addition, IWD will learn personal medical history, food habit, body structure, environmental pollution level, and any other abnormality.

Current ambulance service is just a transporter of patients with limited medical kits. This service should be replaced by H2H services. H2H service will be able to save billion of lives, and thus, it requires immediate implementation on intelligent vehicle. In addition, the mortality rate of accident can be reduced using H2H service. Moreover, intelligent healthcare requires a new business policy for the hospital. Currently, health insurance and direct payment are the business model of hospitals. The health insurance covers only selective diseases which excludes many diseases such as COVID-19, Ebola virus, pregnancy and delivery. The HIH business model will be truly hassle-free treatment without any expenses. But, an intensive research is required for the success of the HIH business model. However, the premium may rise for HIH due to the coverage of every expense related to a person's treatment.

\bibliographystyle{IEEEtran}
\bibliography{mybib}

\begin{IEEEbiography}[{\includegraphics[width=1in,height=1.25in,clip,keepaspectratio]{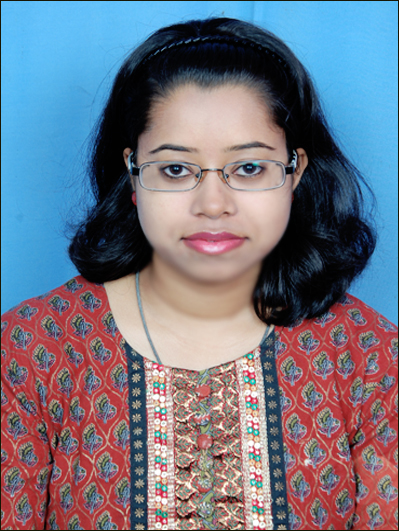}}]{Sabuzima Nayak}(sabuzimanayak@gmail.com) is a research scholar in the Department of Computer Science \& Engineering, National Institute of Technology Silchar, Assam-788010, India. Her research interest is Bioinformatics, Bloom Filter, Communication and Networking, and Big Data. She has published numerous papers in reputed journal, conferences and books.
\end{IEEEbiography}

\begin{IEEEbiography}[{\includegraphics[width=1in,height=1.25in,clip,keepaspectratio]{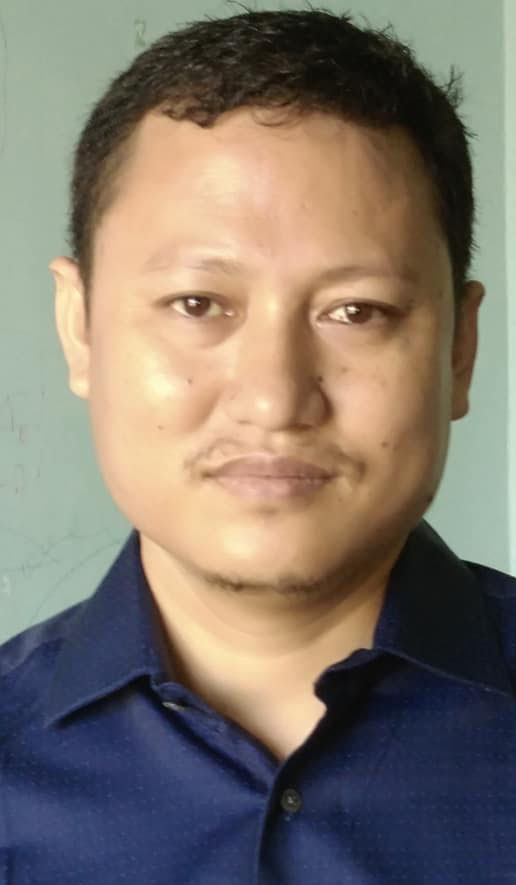}}]{Ripon Patgiri}(ripon@cse.nits.ac.in, patgiri@ieee.org) is currently working as Assistant Professor in the Department of Computer Science \& Engineering, National Institute of Technology Silchar, Assam-788010, India. His research interests are Bloom Filter, Communication and Networks, Big Data, and File System. He has published numerous papers in reputed journal, conferences and books. He is a senior member of IEEE and member of ACM, EAI, and  IETE He has also chaired many conferences.
\end{IEEEbiography}
\vfill

%\vfill


\end{document}